\documentclass[prd,nofootinbib,preprintnumbers,floatfix]{revtex4}
\usepackage[utf8]{inputenc}
\usepackage{amsfonts,amsmath,amssymb}
\usepackage{graphicx}
\usepackage{color}
\usepackage[plainpages=false, colorlinks=true, anchorcolor=blue, linkcolor=blue, citecolor=blue, bookmarks=false]{hyperref}
\usepackage{natbib}
\usepackage{enumitem}
\usepackage{subcaption}
\usepackage[a4paper, left=1.8cm, right=1.8cm, top=2.5cm, bottom=2.5cm]{geometry}

\newcommand{\rthis}[1]{\textcolor{black}{#1}}
\begin{document}
\newcommand{\apjl}{Astrophys. J. Lett.}
\newcommand{\apjs}{Astrophys. J. Suppl. Ser.}
\newcommand{\aap}{Astron. \& Astrophys.}
\newcommand{\aj}{Astron. J.}
\newcommand{\araa}{Ann. Rev. Astron. Astrophys. } 
\newcommand{\aapr}{Astronomy and Astrophysics Review}
\newcommand{\mnras}{Mon. Not. R. Astron. Soc.}
\newcommand{\apss} {Astrophys. and Space Science}
\newcommand{\jcap}{JCAP}
\newcommand{\nar}{New Astronomy Reviews}
\newcommand{\pasj}{PASJ}
\newcommand{\LRR}{Living Reviews in Relativity}
\newcommand{\pasa}{Pub. Astro. Soc. Aust.}
\newcommand{\pasp}{PASP}
\newcommand{\physrep}{Physics Reports}
\newcommand{\ssr}{Space Science Reviews}

\title{ A test of invariance  of halo surface density for FIRE-2 simulations with cold dark matter  and self-interacting dark matter}
\author{ Sujit \surname{K. Dalui}}\altaffiliation{E-mail:ph24mscst11036@iith.ac.in}

\author{Shantanu  \surname{Desai}}  
\altaffiliation{E-mail: shntn05@gmail.com}

\begin{abstract}
Numerous observations have shown that the dark matter halo surface density,  defined as the product of  core radius and halo central density of cored dark matter  haloes is nearly constant and independent of galaxy mass   over a whole slew of galaxy types.  Here we calculate the surface density in cold dark matter (CDM) and self-interacting dark matter (SIDM) models  including baryons, as well as SIDM without baryons,  for dwarf galaxies of masses $\approx 10^{10} M_{\odot}$ using mock catalogs obtained from simulations as part of the Feedback In Realistic Environments  project. We find that the  dark matter surface density and column density are nearly constant for  CDM and SIDM for this mass range. The halo surface density obtained from the Burkert profile fit is consistent with galactic-scale observations within $1\sigma$. We also computed the empirical scaling relations between  the central surface density and maximum velocity using the best-fit dark matter  profiles, and found that they  agree with observations of Milky Way and M31 dwarfs.
\end{abstract}

\affiliation{Department of Physics, Indian Institute of Technology, Hyderabad, Telangana-502284, India}
\maketitle

\section{Introduction}
Over the past few decades, increasingly precise cosmological  observations  have firmly established the $\Lambda$CDM model---comprising approximately 70\% dark energy and 25\% cold dark matter---as the standard cosmological paradigm. While it successfully explains the large-scale structure~\citep{Planck18}, a few tensions have crept up in this standard model. Some of these problems include Hubble tension\citep{Divalentino}, failure to detect cold dark matter candidates in laboratory based experiments~\cite{Merritt}, CMB anomalies at large angular scales~\cite{Copi}, core-cusp and the  missing satellites problem~\cite{Bullock,Morgan}, Lithium-7 problem in Big-Bang nucleosynthesis~\cite{Fields}, radial acceleration relation in spiral galaxies with very low scatter~\citep{McGaugh16}, etc. An up-to-date  summary of  some of the challenges for $\Lambda$CDM model, and possible alternatives have been recently reviewed in a number of works~\cite{Periv,Abdalla22,Peebles22,alternatives,Banik,Cosmoverse}.


One such anomaly is the observational evidence  showing   that the central surface density of a dark matter halo  is constant over a wide range of systems spanning over 18 orders of magnitude in blue-band luminosity, and including various  galaxy types such as spirals, low surface brightness galaxies, and dwarf spheroidal satellites of the Milky Way~\cite{Kormendy04,Donato,kormeddy,Salucci19}. The halo surface density is defined as follows:
\begin{equation}
 S \equiv \rho_c \times r_c, 
\label{eq:halosurfacedensity}
\end{equation}
where $r_c$ is the core radius and $\rho_c$ is the central density for a cored dark matter density profile such Burkert or the isothermal profile. The halo surface density  is also proportional to  the Newtonian acceleration at the radius $r_c$~\cite{Gentile09,Cardonedel}, and hence the constancy of halo surface density can also be interpreted as constancy of Newtonian acceleration at the core radius~\cite{Gentile09}.
The most recent best-fit value, obtained by fitting all available data for the systems above, is given by $\log(\rho_c r_c) = 2.15 \pm 0.2$, with units of $\log(M_{\odot},\mathrm{pc}^{-2})$~\cite{Salucci19}.

Since not all dark-matter-dominated systems can be fitted by a cored profile, a variant of the aforementioned halo surface density has also been proposed in the literature called dark-matter  column density which can be evaluated for any density profile $\rho(r)$ as follows~\cite{Boyarsky,DelPopolo12}:
\begin{equation}
S^* (R)  = \frac{2}{R^2} \int_0^{R} r' dr' \int_{-\infty}^{+\infty} dz \rho_{DM} (\sqrt{r'^2+z^2})
\label{eq:SB}
\end{equation}
The numerical value of the column density is also close to the surface density depending on the choice of radius at which it is evaluated~\cite{DelPopolo12}.
However, the constancy of the dark matter halo surface density or column density has been disputed in a number of works, which have found correlations between the halo surface density and the galaxy luminosity, the halo mass, the stellar age, and the surface brightness~\cite{Boyarsky,Napolitano,Cardone, DelPopolo12,Cardonedel,Saburova,DelPopolo17,DelPopolo20,Del23}. 
Some of these  works have found a power-law scaling with halo mass,  $S \propto M_{halo}^{0.16-0.2}$~\citep{DelPopolo12},  and also with  luminosity, $S \propto L^{0.13}$~\citep{DelPopolo20}.
Observationally, it has been shown that the constancy of dark matter surface halo density does not extend to galaxy group~\cite{Gopika2021} and galaxy cluster scales~\cite{Chan,Gopika_2020}. Therefore the dark matter halo surface/column density cannot be a universal constant for all dark matter dominated systems.

Nevertheless, these observations of   (near)-constancy of the halo surface density have been used as a laboratory  for testing alternatives to $\Lambda$CDM~\cite{Milgrom09,Berezhiani,Kaplinghat,Lin,Burkert20,Chavanis}. These applications have recently been reviewed in ~\cite{Gopika23}. In particular,  it has been shown that the observed dark matter halo surface density  is in tension with predictions  from fuzzy dark matter models~\cite{Burkert20}.
The first systematic test of the constancy of the halo surface density using mock galaxy catalogs  from $\Lambda$CDM simulations  in the mass range from $\sim 10^{11} M_{\odot} - 10^{13} M_{\odot}$   was carried out in ~\cite{Gopika23}. This work considered
multi-wavelength mock galaxy catalogs from $\Lambda$CDM simulations, where the adiabatic contraction of dark matter halos in the presence of baryons was taken into account~\cite{mock}. This work found that  after fitting the mock catalogs to a generalized NFW profile, the halo surface density (after averaging over concentration) is consistent with a flat surface density~\cite{Gopika23}. However, because of the intrinsic scatter, a power-law dependence as a function of halo mass could also not be ruled out~\cite{Gopika23}. Most recently a connection between the scaling relations for halo surface density and  mass-concentration ($c-M$) relations from cosmological N-body simulations has been discussed in ~\citet{Kaneda} (K24 hereafter). (See also ~\cite{Almeida}). This work was able to reproduce the observational results  on the halo surface density over a diverse range of dark matter systems from  dwarf satellites to galaxy clusters, using the $c-M$ relation from the Uchuu suite of simulations~\cite{Moline23}.
K24 also proposed a new scaling relation that takes into account the transformation from cusps to cores, which they argued happens around $10^{11} M_{\odot}$~\cite{Kaneda}. K24 subsequently derived a relation between the maximum circular velocity and the mean halo surface density and compared with data from Milky way and M31 dwarfs. Most recently, it was shown that the near-constancy of the halo surface density holds for any generic dark-matter model (beyond $\Lambda$CDM) as well as any baryonic feedback process that  redistributes dark matter within the haloes~\cite{Almeida}.

In this work, we carry out a test of the halo surface density  (along the same lines as ~\cite{Gopika23}) using the properties of dark matter haloes 
for  dwarf galaxy mass scales ($M_{halo} \approx 10^{10} M_{\odot}$) at redshift $z=0$. These dark matter haloes have been obtained using cosmological N-body simulations, which are a part of the  Feedback in Realistic Environments (FIRE-2) project, and  have recently been studied in detail in ~\citet{Maria_C} (S25, hereafter).  These simulations consist of both  $\Lambda$CDM and Self-Interacting Dark Matter and also incorporate baryonic feedback processes~\cite{Maria_C}. Although the dynamic range of the halo mass  is not large enough to probe a mass dependence of the halo surface density, it would still be interesting to check if the surface density from these simulations agrees with observational data. This is the main motivation behind this work.

This manuscript is organized as follows. In Sec.~\ref{sec1} we provide a brief description of the FIRE-2 simulation models used for calculating surface density in S25. In Sec.~\ref{sec2},  we  outline  the procedure for  estimating the  parameters of  the different dark matter profiles. In Sec.~\ref{sec3} we discuss  the results for the dark matter halo surface and column density. In Sec.~\ref{sec:c-M}, we estimate the relation between maximum circular velocity and mean surface density using the dark matter profiles and compare them to observations of dwarf satellites.
Finally, we conclude in Sec.~\ref{sec:last}.

\section{Simulations used}
\label{sec1}

For this work, we used the dark matter density profiles from FIRE-2 simulations\footnote{\url{http://fire.northwestern.edu}}(for both CDM and SIDM) \cite{FIRE_Hopkins,FIRE_Fitts,FIRE_Robles} with complete baryonic physics for eight dwarf galaxies with increasing stellar mass, as outlined in S25.  For SIDM, there is also a corresponding set of simulations using only dark matter (labelled as SIDM DMO). Here, we briefly summarize some of the salient features of these simulations.  More details of the underlying FIRE-2 simulation and the full baryonic 
physics  employed  can be found in S25.

The simulations consist of isolated haloes with virial masses of $10^{10} M_{\odot}$ with a variation of $\pm 30\%$ and stellar masses of 
$M_\star \approx 10^5 - 10^7\, M_\odot$. These haloes were labelled in S25 as m10b, m10c, and so on up to m10k. We uadopt the same nomenclature as S25 for designating  these haloes.
 Each CDM simulation has an analogous SIDM version with identical initial conditions and identical physics along with a self-interaction cross section $\sigma / m = 1\ \mathrm{cm}^2\, \mathrm{g}^{-1}$~\cite{rocha_cosmological_2013}.  The SIDM simulations have also been presented for one halo with $\sigma/m$ of 0.1 $cm^2/g$ and 10 $cm^2/g$. However, we have not analyzed the profiles from these two simulations, since with one halo it is not possible to probe the variation of the halo surface density.
 The baryonic physics and feedback processes used in the simulations  are implemented using the \textsc{GIZMO} simulations~\cite{hopkins_gizmo_2015}.  Both the CDM and SIDM simulations also have a dark matter only component without any feedback processes implemented. The dark matter only simulations also have the same initial conditions as their baryonic counterpart, except for the dark matter component replacing the baryons. Because of this, the particle masses are elevated by a factor of 1.2 for the dark matter only simulations. 


From the simulations, dark-matter haloes with a viral radius enclosing a region of average density $\Delta_{\rm vir}(z)\rho_{\rm crit}(z)$, where $\Delta_{\rm vir}(z)$ is the redshift-dependent virial overdensity~\cite{bryan_statistical_1998} and $\rho_{\rm crit} = 3H^2(z)/8\pi G$, are found using the ROCKSTAR halo finder~\cite{behroozi_rockstar_2012}. The dark matter density profiles are constructed  using 25 logarithmically spaced radial bins, starting from the convergence radius and extending to each halo's virial radius.
The data used for the analysis can be found in Table A1 of S25 . The density profiles have been kindly shared with us by the authors of S25 (Maria Straight, private communication).

\section{Density profiles with fitted parameters}
\label{sec2}
We now outline the data analysis procedure used to calculate  the halo surface density. First, we describe the various parametric models employed to characterize the dark matter halo density profiles. These models are used to fit the simulation data for  CDM+baryons, SIDM+baryons, and SIDM DMO  scenarios. This is followed by a description of the regression technique employed to obtain the best-fit parameters for each of the profiles. 

\subsection{Dark Matter Density Profiles}
S25 has tabulated the  best-fit parameters  for core-Einasto and  $\alpha\beta\gamma$ profile fits to SIDM and CDM simulations with baryons, and we directly use their fitted values for calculating the dark matter column density.
However, since the SIDM DMO fits  have not been provided in S25, we fit them to core-Einasto and $\alpha\beta\gamma$ profile profiles. These best-fit values can be found in Appendix~\ref{sec:appendix}.
For all the three simulations,  we independently fit the density  to Burkert profile, allowing for a direct comparison with previous results in literature on  halo surface  density obtained with Burkert profile. We now discuss the parametric forms of  the dark matter density profiles for each model.
\begin{itemize}
\item Burkert Profile~\cite{Burkert95}. This is the most widely used cored profile and can be written as follows:
 \begin{equation}
 \rho_{Bur}(r) =\frac{\rho_c r_c^{3}}{(r^2+r_c^2)(r+r_c)},
 \label{eq:eqbur} 
 \end{equation}
where $r_c$ and $\rho_c$ denote the core radius and core density, respectively.

\item Core-Einasto Profile~\cite{lazar_dark_2020}
 \begin{equation}
 \rho_{\mathrm{cEin}}(r) = \tilde{\rho}_{\rm s} \exp \left\{-\frac{2}{\hat{\alpha}}\left[\left(\frac{r+r_{\rm c}}{\tilde{r}_{\rm s}}\right)^{\hat{\alpha}}-1\right]\right\}.
 \label{eqn:coredEinasto}
 \end{equation}
\end{itemize}
The core-Einasto profile is a cored version of the original Einasto profile~\cite{einasto_construction_1965}, which was introduced to account for baryonic feedback in CDM simulations~\cite{lazar_dark_2020}.
 In Eq.~\ref{eqn:coredEinasto}, $r_{\rm c}$ is the dark matter core radius and $\tilde{r}_{\rm s}$ and $\tilde{\rho}_{\rm s}$ are the  radius and density free parameters. 
\begin{itemize}
\item $\alpha\beta\gamma$-profile~\cite{zhao_analytical_1996}
 \begin{equation}
    \rho_{\alpha\beta\gamma}(r) = \dfrac{\rho_{\rm s}}{(r/r_{\rm s})^{\gamma_{\rm s}}\left[1+(r/r_{\rm s})^{\alpha_{\rm s}}\right]^{(\beta_{\rm s}-\gamma_{\rm s})/\alpha_{\rm s}}} . 
 \label{eqn:alphabetagamma}
 \end{equation}
\end{itemize}
where $r_{\rm s}$ and $\rho_{\rm s}$ denote the scale radius and scale density, and the inner and outer slopes are parameterized by $\gamma_{\rm s}$ and $\beta_{\rm s}$, with $\alpha_{\rm s}$ setting the transition rate between them. This profile is also known in literature  as the generalized NFW  (gNFW) profile. 

\subsection{Fitting Procedure}
Since no uncertainties  have been provided for  the dark matter profile at a given radius , we cannot find the best-fit parameters by  $\chi^2$ minimization or obtain the uncertainties in the best-fit parameters.  Therefore, similar to S25, we find the  best-fit parameters (for each of the aforementioned profiles)  by minimizing the figure-of-merit function ($Q^2$), which is defined as follows~\cite{navarro_diversity_2010}:
\begin{equation}
    Q^2 = \frac{1}{N_{\rm bins}}\sum_{i=1}^{N_{\rm bins}}[\ln{\rho(r_i)}-\ln{\rho_{\rm model}(r_i)}]^2
\label{eqn:quality}
\end{equation}
Although it is difficult to quantify a threshold $Q$ value (unlike reduced $\chi^2$) below which fits are reliable~\cite{navarro_diversity_2010},
  we have ignored the fitted profiles with $Q$ values higher than 0.25 in the calculation of surface density. This threshold $Q$ value  was  based on a visual inspection of the fitted profiles.

\begin{figure}[http]
\centering
\includegraphics[width=0.6\linewidth]{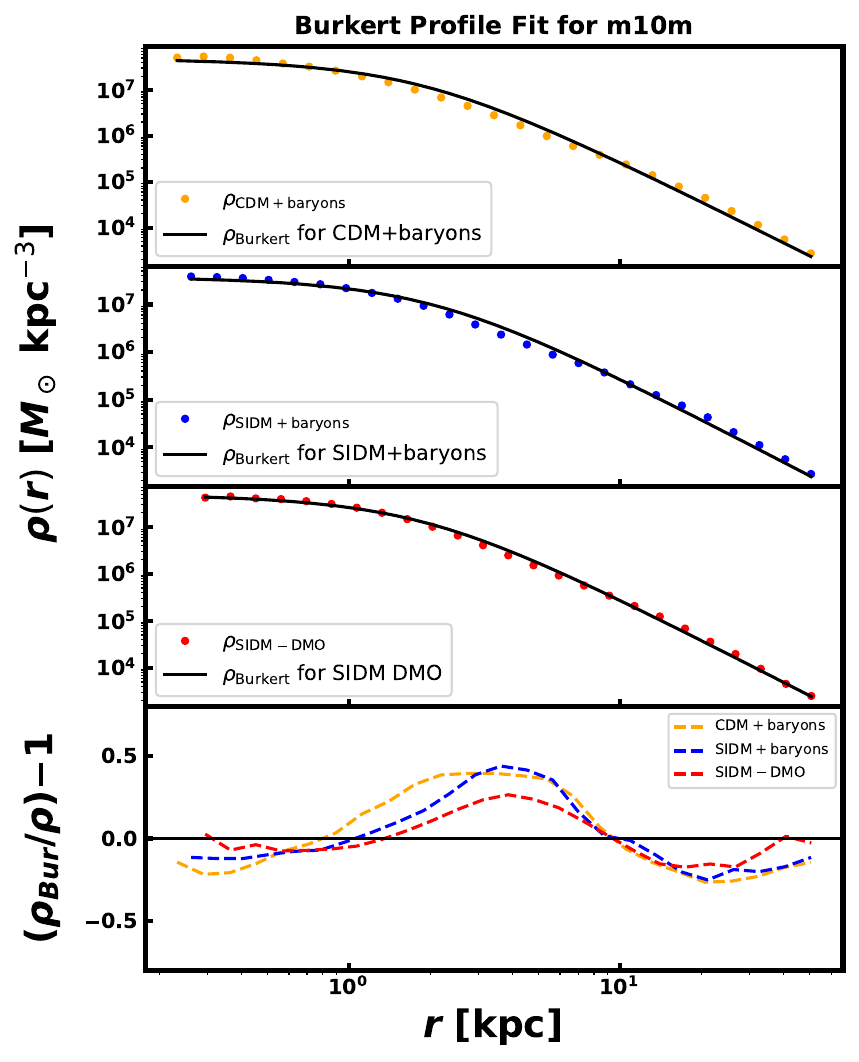}
\caption[]{ This figures shows the density profiles along with the  fitted Burkert profile for the CDM+baryons, SIDM+baryons, and SIDM DMO of halo m10m, simulated with full galaxy formation physics. Here, the solid lines are the fitted Burkert profiles and dotted lines indicate the  density distribution. \rthis{We have shown the residuals for above 3 plots. The residuals show that the fits are similar and the maximum value around r $\sim$ 4 kpc.}}
\label{fig:bucket_fit}
\end{figure}

 The best-fit parameters for the Burkert  profiles for all the three simulations have been tabulated in Table \ref{tab:fitB}. In Figure \ref{fig:bucket_fit} we have shown the density profiles and the fitted Burkert profile for the CDM, SIDM,  and SIDM DMO  for one representative halo (m10m) simulated with full galaxy formation physics. 
 For CDM simulations, the profiles  with lower stellar mass $\sim (4.6\times10^5$ -- $7.8\times10^6$ $M_\odot$) have  cuspy dark matter profiles, and hence  the Burkert profile does not provide a good fit to those distributions with  Q values $>0.25$ as shown in Table \ref{tab:fitB}. We have excluded m10b, m10c, m10d, m10e, m10f and m10h (Q $>0.25$) for our calculation of surface density ($S_{Bur}$) and column density($S^*_{Bur}$).
 Therefore, for CDM profiles we were able to show the dark matter surface density profile for only two masses.

For the core-Einasto profiles, $\hat{\alpha}$ is assumed to be constant with a value of 0.16~\cite{gao_redshift_2008}, while the best-fit values  for the  remaining three parameters in Eq.~\ref{eqn:coredEinasto} can be found in Table A1 of S25 for CDM and SIDM haloes with baryons.
Similarly for the $\alpha\beta\gamma$-profile, $\beta_{\rm s}=2.5$ and $\gamma_{\rm s} = 0$ are kept fixed~\cite{di_cintio_mass-dependent_2014} similar to S25,
while the best-fit values for the remaining free parameters can be found in Table A1 of S25.
Finally, the best-fit values for core-Einasto and $\alpha\beta\gamma$-profile for SIDM DMO simulations are tabulated in Table~\ref{tab:g-E_fit_DMO}.


\begin{table}[h]
\centering
\caption{Best-fit parameters for Burkert profile. We note that
$M_{200}$ is calculated from density distribution at virial radius $r_{200}$. (We have directly used the density profile data points  for calculating $M_{200}$, and not used any of the dark matter fits.)}
\begin{tabular}{lcccc}
    \hline
    $\begin{array}{c} 
        CDM\ halo \\ (+baryons)
        \end{array} $& 
        $ \begin{array}{c} 
        M_{200} \\ (M_\odot)
        \end{array} $ &
        $ \begin{array}{c} 
        r_c \\ (kpc)
        \end{array} $ &
        $ \begin{array}{c} 
        \rho_c \\ (M_\odot \, kpc^{-3})
        \end{array} $ &
        $ \begin{array}{c} 
        Q_{burkert}\\
        \end{array} $\\
    \hline
    m10b & $5.0\times10^9$ & 1.47 & $6.60\times10^7$ & 0.498\\
    m10c & $4.5\times10^9$ & 1.76 & $4.08\times10^7$ & 0.465\\
    m10d & $4.6\times10^9$ & 1.39 & $7.20\times10^7$ & 0.477\\
    m10e & $5.1\times10^9$ & 2.07 & $2.94\times10^7$ & 0.433\\
    m10f & $5.2\times10^9$ & 1.34 & $8.55\times10^7$ & 0.340\\
    m10h & $7.0\times10^9$ & 1.67 & $6.56\times10^7$ & 0.366\\
    m10k & $6.7\times10^9$ & 2.04 & $3.79\times10^7$ & 0.227\\
    m10m & $7.3\times10^9$ & 1.85 & $4.98\times10^7$ & 0.227\\
    \hline
    \multicolumn{1}{l}{SIDM halo+baryons} \\
    \hline
    m10b & $5.1\times10^9$ & 1.70 & $4.75\times10^7$ & 0.395\\
    m10c & $4.5\times10^9$ & 1.97 & $3.11\times10^7$ & 0.366\\
    m10d & $4.8\times10^9$ & 1.71 & $4.40\times10^7$ & 0.294\\
    m10e & $5.0\times10^9$ & 2.54 & $1.80\times10^7$ & 0.332\\
    m10f & $5.1\times10^9$ & 1.52 & $6.26\times10^7$ & 0.248\\
    m10h & $7.0\times10^9$ & 2.02 & $4.16\times10^7$ & 0.246\\
    m10k & $6.6\times10^9$ & 2.33 & $2.73\times10^7$ & 0.125\\
    m10m & $7.3\times10^9$ & 2.04 & $3.91\times10^7$ & 0.193\\
    \hline
    \multicolumn{1}{l}{SIDM DMO halo} \\
    \hline
    m10b & $6.3\times10^9$ & 1.92 & $3.62\times10^7$ & 0.361\\
    m10c & $5.6\times10^9$ & 2.08 & $2.88\times10^7$ & 0.294\\
    m10d & $5.3\times10^9$ & 1.76 & $4.31\times10^7$ & 0.225\\
    m10e & $5.6\times10^9$ & 1.99 & $3.39\times10^7$ & 0.383\\
    m10f & $5.4\times10^9$ & 1.52 & $6.71\times10^7$ & 0.216\\
    m10h & $7.6\times10^9$ & 1.90 & $5.00\times10^7$ & 0.182\\
    m10k & $7.1\times10^9$ & 2.00 & $4.24\times10^7$ & 0.168\\
    m10m & $7.7\times10^9$ & 1.86 & $5.14\times10^7$ & 0.125\\
    \hline
\end{tabular}
\label{tab:fitB}
\end{table}

\section{Results for dark matter halo surface  and column density}
\label{sec3}

We now calculate the dark matter column density and surface density using the best-fit parameters for all the dark matter profiles for the three sets of simulations. We note that the dark matter column density depends on the radius up to which it is evaluated. Most of the observational studies on dark matter halo surface density have been obtained by assuming a Burkert or similar isothermal profiles~\cite{Spano08,Donato,kormeddy,Salucci19,Chan,Gopika_2020,Gopika2021}. The dark matter column density has only been obtained for NFW profile~\cite{Boyarsky} and Burkert profile at galaxy group scales~\cite{Gopika2021}. 
Therefore, it is not possible to directly compare the column density obtained using  any of the three profiles we have considered in this work  with observations of dwarf galaxies.
Therefore, we can only compare the halo surface density obtained using the best-fit Burkert profile with  the observational fit of $\rho_c r_c = (141 \pm 65) M_{\odot} pc^{-2}$~\cite{Salucci19}. For all profiles (including Burkert), we show the column density as a function of halo mass, mainly to test the constancy of dark matter surface density.

We now obtain the expression for dark matter column density
by plugging in the expression for dark matter density profiles in Eq.~\ref{eq:SB}.
The dark matter column density for the Burkert profile can written as follows:
\begin{equation}
\begin{split}
S^*_{\mathrm{Bur}} (R) = \frac{2}{R^2} \int_0^{R} r' \, dr' 
\int_{-\infty}^{+\infty} dz \,\frac{\rho_c r_c^{3}}{(r_d^2+r_c^2)(r_d+r_c)}  \\
\text{where} \quad r_d = \sqrt{r'^2 + z^2}
\end{split}
\label{eq:SBUR1}
\end{equation}
For the Burkert profile, we calculate $S^*_{Bur}$  at $R = 1.66r_c$ similar to ~\cite{Gopika2021}.

Similarly for the core-Einasto profile, the column density can be written as follows:
\begin{equation}
\label{eq:cEin}
\begin{split}
S^*_{\mathrm{cEinasto}} (R) = \frac{2}{R^2} \int_0^{R} r' \, dr' 
\int_{-\infty}^{+\infty} dz \,\tilde{\rho}_{\rm s} \exp \left\{-\frac{2}{\hat{\alpha}}\left[\left(\frac{r_d+r_{\rm c}}{\tilde{r}_{\rm s}}\right)^{\hat{\alpha}}-1\right]\right\}  \\
\end{split}
\end{equation}
where  \( r_d = \sqrt{r'^2 + z^2} \). For the core-Einasto profile, we evaluate $S^*_{\mathrm{cEinasto}}$ at $R = \tilde{r}_s$.

Finally, for the $\alpha\beta\gamma$ profile, the column density can be written as:
\begin{equation}
\label{eq:gnfw}
\begin{split}
S^*_{\mathrm{\alpha\beta\gamma}} (R) = \frac{2}{R^2} \int_0^{R} r' \, dr' 
\int_{-\infty}^{+\infty} dz \,\dfrac{\rho_{\rm s}}{
\left( \dfrac{r}{r_{\rm s}} \right)^{\gamma_{\rm s}} 
\left[ 1 + \left( \dfrac{r}{r_{\rm s}} \right)^{\alpha_{\rm s}} \right]^{\frac{\beta_{\rm s} - \gamma_{\rm s}}{\alpha_{\rm s}}} 
}
\end{split}
\end{equation}
\text{where} \quad \( r_d = \sqrt{r'^2 + z^2} \). 
For the Burkert profile, we calculate the  column density ($S^*_{Bur}$)  using Eq.~\ref{eq:SBUR1} at $R = 1.66r_c$ similar to ~\cite{Gopika2021}.
Similarly, for the core-Einasto profile using Eq.~\ref{eq:cEin} at R = $\tilde{r}_s$.
Finally, the column density for $\alpha\beta\gamma$ profile was calculated using Eq.~\ref{eq:gnfw} at $R=r_s$ similar to our analysis  using mock $\Lambda$CDM catalogs~\cite{Gopika23}.

\begin{figure}[htbp]
    \centering
    \begin{minipage}[t]{0.48\linewidth}
        \centering
        \includegraphics[width=\linewidth]{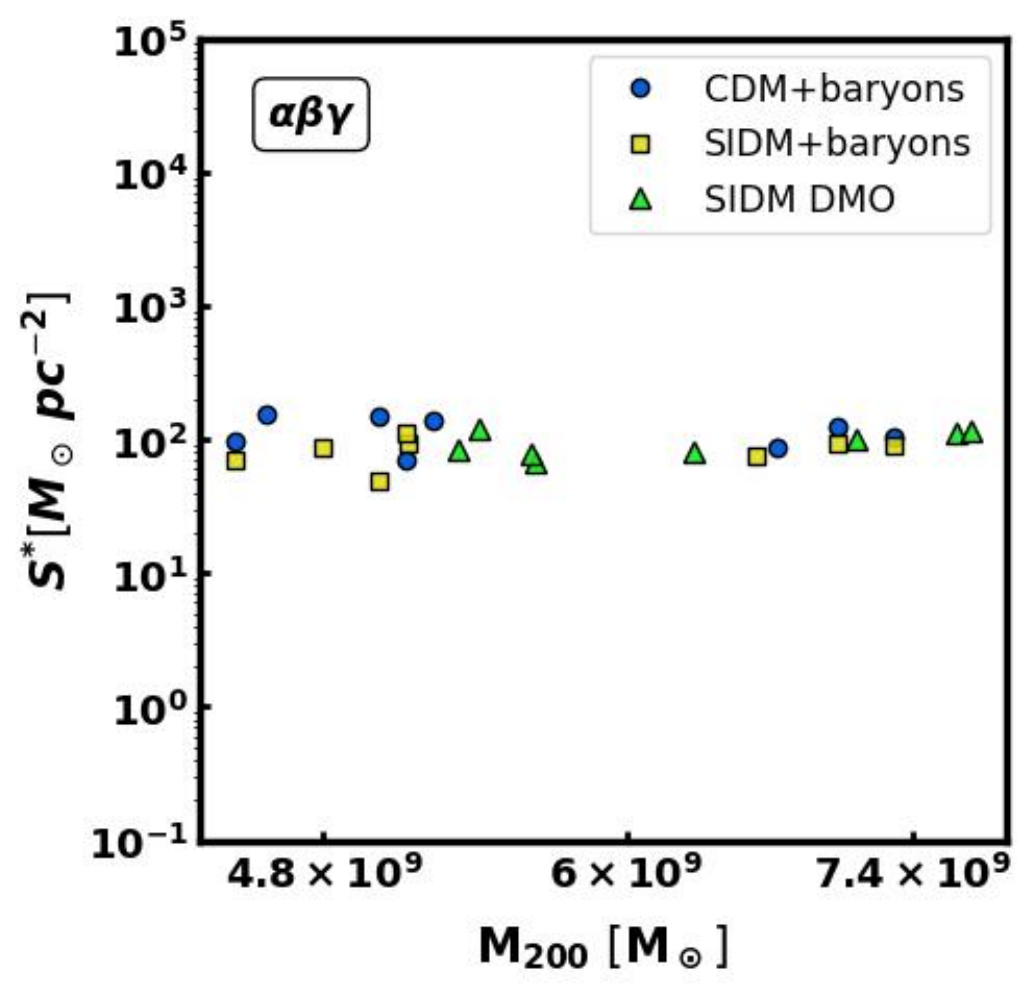}
        \caption{ The column density for $\alpha\beta\gamma$ profile
        as a function of halo mass for all the three sets of simulations (CDM+baryons, SIDM+baryons, SIDM DMO only) evaluated using Eq.~\ref{eq:gnfw} at $r_{200}$. We find that the column density is roughly constant with over this mass range for all the three simulations.}
        \label{fig:g_column}
    \end{minipage}
    \hfill
    \begin{minipage}[t]{0.48\linewidth}
        \centering
        \includegraphics[width=\linewidth]{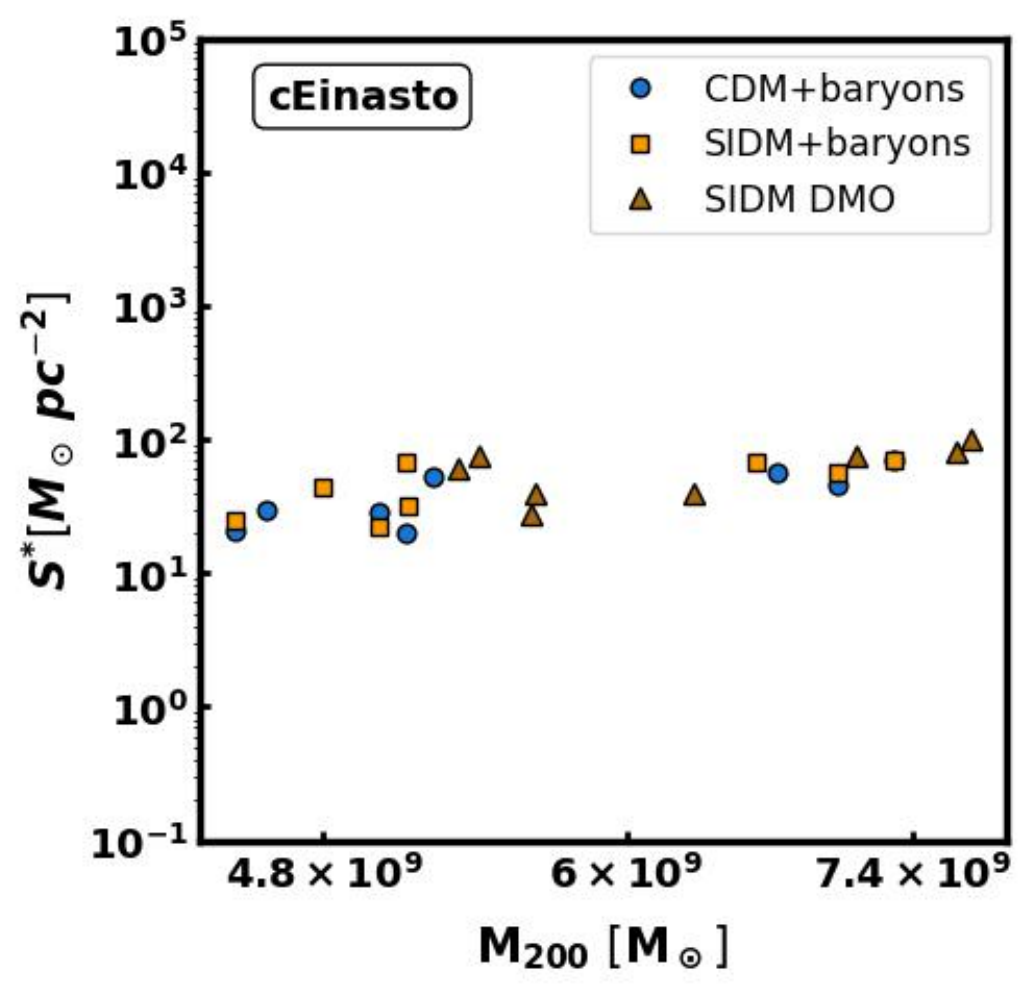}
        \caption{ The  column density ($S^*_{\mathrm{cEin}}$) for core-Einasto profile evaluated using Eq.\ref{eq:cEin} at $R =\tilde{r}_s$ for the same set of simulations as in Fig.~\ref{fig:g_column}, as a function of $M_{halo}$. Once again, we find that the column density is roughly constant with  for  over this mass range for all the three simulations.}
        \label{fig:cE_column}
    \end{minipage}
    
    \vspace{1cm}
    
    \begin{minipage}[t]{0.48\linewidth}
        \centering
        \includegraphics[width=\linewidth]{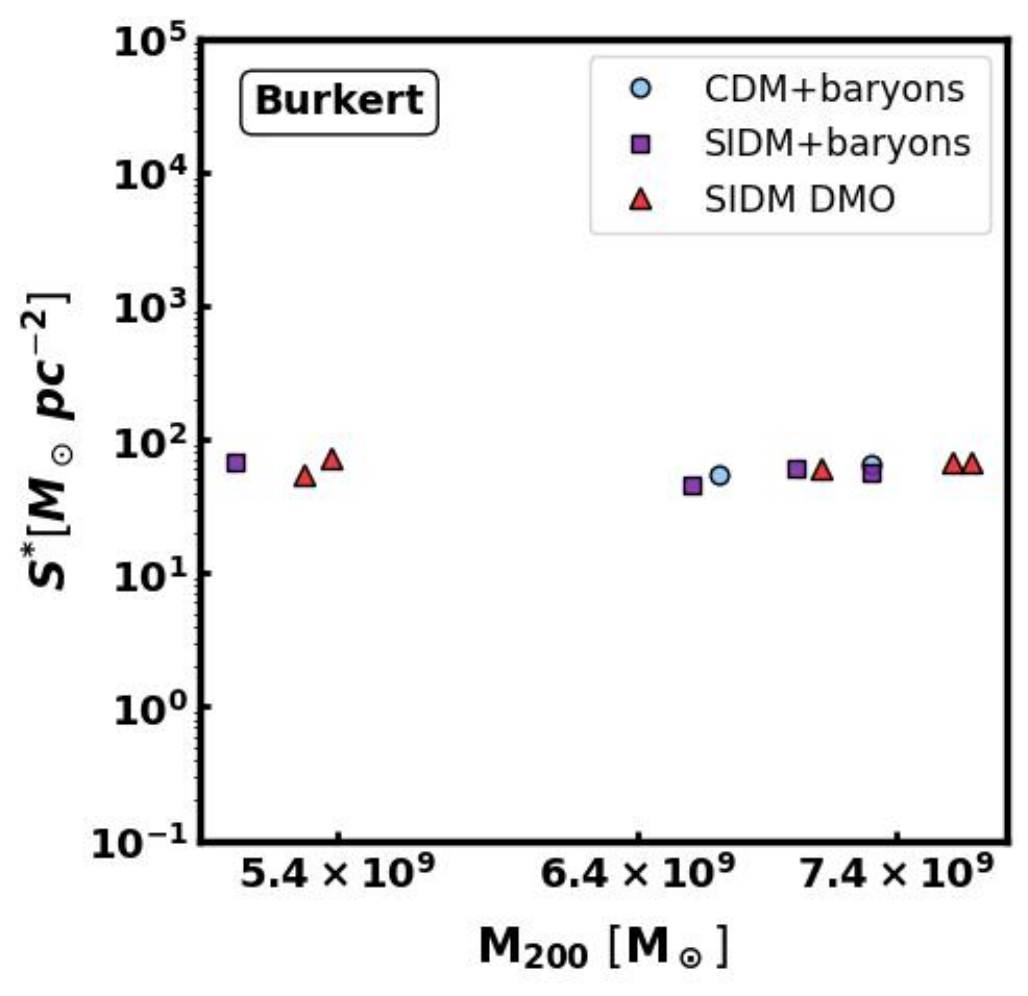}
        \caption{ The column density ($S^*_{Bur}$) evaluated using Eq.~\ref{eq:SBUR1} at $R = 1.66r_c$ for the same set of simulations as in Fig.~\ref{fig:g_column}, as a function of $M_{halo}$.  Once again, it is roughly constant. We note that density profiles with $Q>0.25$ have been removed from this analysis. The haloes which have been included in the figure include m10k and m10m for CDM+hydro, m10f, m10h, m10k and m10m for SIDM+hydro and m10d, m10f, m10h, m10k and m10m.}
        \label{fig:bur_column}
    \end{minipage}
    \hfill
    \begin{minipage}[t]{0.48\linewidth}
        \centering
        \includegraphics[width=\linewidth]{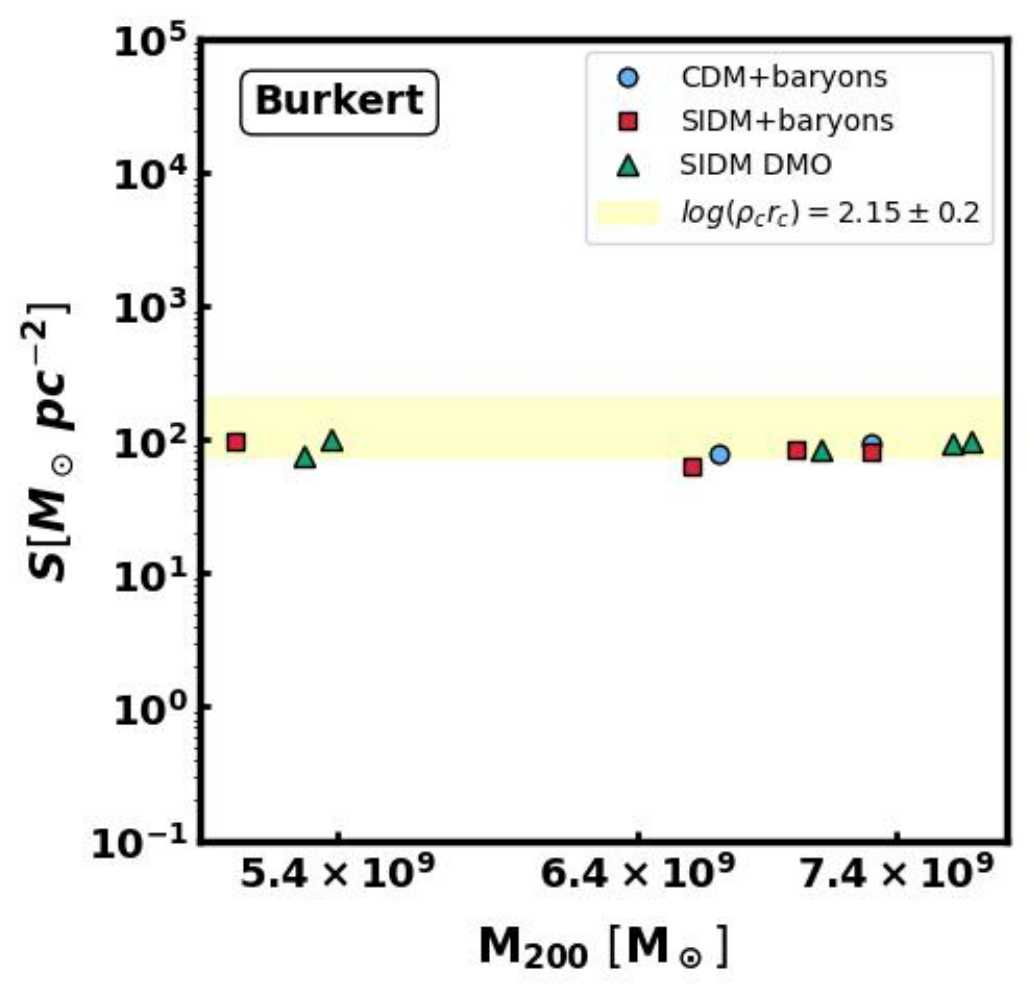}
        \caption{Here surface density($S_{Bur}$) is calculated using Eq.~\ref{eq:halosurfacedensity} where $\rho_c$ and $r_c$ are scale parameter of fitted Burkert profile with Q values less than 0.25. $S_{Bur}$ gives a constant surface density with $M_{halo}$. We have also overlayed $S_{Bur}$ with $\rho_c r_c = (141 \pm 65) M_{\odot} pc^{-2}$~\cite{Salucci19}.}
        \label{fig:bur_surf}
    \end{minipage}
\end{figure}

The plots for the column density  as a function of halo mass (using $M_{200}$) can be found in Figs.~\ref{fig:g_column}, \ref{fig:cE_column}, \ref{fig:bur_column}  for $\alpha\beta\gamma$, core-Einasto, and Burkert profile, respectively. 
The plot for the halo surface density (with Burkert profile) can be found in  Fig.~\ref{fig:bur_surf}. 
We  find that the column density  is  nearly constant irrespective of the density profile we used for all the three simulations used.  Same is the case for the halo surface density.
We then calculate the median of  the surface density for all the three simulations  along with the  1$\sigma$ uncertainties for all haloes having $Q<0.25$, where the $1\sigma$ uncertainties  were obtained from the 16th and 84th percentiles.
Since we have only two datasets (m10k and m10m) for CDM+hydro, we did not calculate the uncertainties explicitly for this simulation. We have tabulated all these values for all combinations of dark matter profiles and simulations in Table~\ref{tab:surface_density}. We find that the  results for the halo surface density for all the three simulations  are  consistent  with the observational estimate of $\rho_c r_c = (141 \pm 65) M_{\odot} pc^{-2} $~\cite{Salucci19} to within $1\sigma$. The column densities for other profiles are also close to $S_{Bur}$. \rthis{We also note that the values $S^*_{\mathrm{\alpha\beta\gamma}}$ in this mass range ($\sim 10^{9} M_{\odot}$) are consistent with those found in ~\cite{Gopika23} (for the mass range of $10^{11}-10^{13} M_{\odot}$).}

Therefore, our results for the halo surface density for all the three simulations  are  consistent within $1\sigma$ with the observational estimate of $\rho_c r_c = (141 \pm 65) M_{\odot} pc^{-2} $~\cite{Salucci19}.

\begin{table}[h]
\caption{Dark matter column density and surface density (for Burkert profile) for each of the three simulations using the best-fit dark matter profiles. Each row shows the Median and the 1$\sigma$ uncertainties for the eight  haloes in each simulation for which
$Q<0.25$. For CDM+baryons haloes, we could only get two haloes with $Q<0.25$ while fitting to Burkert profiles.  Therefore, we  have only shown the mean values without any error bars for CDM+baryons haloes fitted using Burkert profile.}
\centering
\begin{tabular}{lcccc}
    \hline
     &
    $\begin{array}{c} 
        S^*_{\mathrm{\alpha\beta\gamma}}\\( M_\odot \, pc^{-2})\\(1)
        \end{array} $ &
    $\begin{array}{c}
        S^*_{\mathrm{cEinasto}}\\( M_\odot \, pc^{-2})\\(2)
        \end{array} $ &
    $\begin{array}{c}
        S^*_{\mathrm{Bur}}\\( M_\odot \, pc^{-2})\\(3)
        \end{array} $ &
    $\begin{array}{c}
        S_{Bur}(\rho_c r_c)\\( M_\odot \, pc^{-2})\\(4)
        \end{array} $\\
    \hline
    CDM halo+baryons & $114^{+34}_{-27}$ & $37^{+19}_{-16}$ & 60 & 85\\
    SIDM halo+baryons& $89^{+5}_{-18}$ & $50^{+17}_{-24}$ & $58^{+6}_{-8}$&$82_{-10}^{+8}$\\
    SIDM DMO halo & $92^{+22}_{-13}$&$67^{+12}_{-28}$&$68^{+2}_{-9}$&$95_{-13}^{+3}$\\
    \hline
    \end{tabular}
\label{tab:surface_density}
\end{table}

\section{Comparison with K24}
\label{sec:c-M}
Most recently, K24  has found a correspondence between the dark matter halo surface/column density scaling relations and results from cosmological N-body simulations. They considered two models for dark matter haloes corresponding to cuspy and cored haloes.
For cuspy haloes, K24 have  shown that the results for constant surface density can be reproduced using NFW profile with   concentration-mass ($c-m)$ relation~\cite{Moline23} obtained from the Uchuu suite of cosmological simulations~\cite{Ishiyama} for all dark matter dominated systems, except for dwarf galaxies. Then, K24  considered a cusp to core transformation model based on the mechanism proposed in ~\cite{Ogiya14}. They  equated the parameters of the NFW profile with a Burkert profile, and subsequently re-derived the surface/column density using the transformed Burkert parameters. 


K24 used both  the cuspy and cored dark matter parameters  to calculate the  mean surface density of the dark matter halo (${\Sigma}(< r_{max})$)  within the radius corresponding to  the  maximum circular velocity ($V_{max}$).
\begin{equation}
      \bar{\Sigma}(<r_{max})=\frac{M (<r_{max})}{\pi r_{max}^2}, \label{eq:Sigma_vmax_definition}
\end{equation}
where $M(<r_{max})$ is the total dark matter mass, $r_{max}$ is the radius corresponding to $V_{max}$.
\begin{equation}
M(<r_{max})=\int_0^{r_{max}}4\pi\rho_\mathrm{DM}\left(r'\right)r'^2dr',
\end{equation}
where $\rho_\mathrm{DM}$ is the assumed dark matter density distribution.
$V_{max}$ can be obtained by maximizing the circular velocity $V(r)$ given by: 
\begin{equation}
   V(r) = \sqrt {\frac{G M(<r)}{r}}
\end{equation}
K24 then  constructed a relation  between $\Sigma(< r_{max})$  and $V_{max}$, and compared them with observations of Milky Way and M31 dwarfs from the compilation in ~\cite{Hayashi17}.

\begin{figure*}[htbp]
\centering
\includegraphics[width=0.6\linewidth]{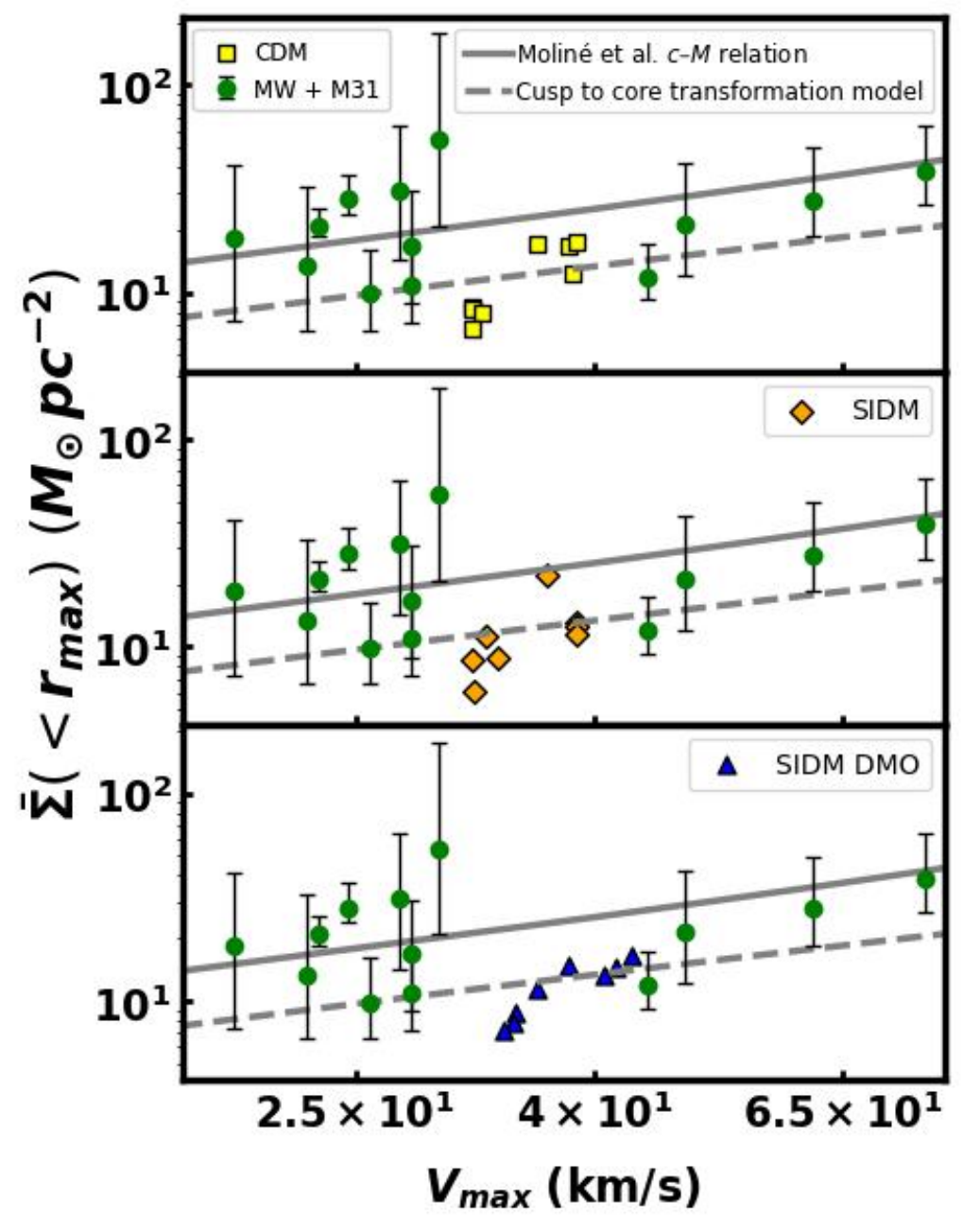}
\caption{This plot shows the relation between $\bar{\Sigma}(<r_{max})$ and $V_{max}$ for all the three cosmologies (CDM+baryons, SIDM+baryons,   and SIDM DMO) as described in the legend. These relations are compared with $c$--$M$ relation~\cite{Moline23} considering a cuspy density profile (solid line) and cusp to core transformation model (dashed line)~\cite{Kaneda}.  We have also shown $\bar{\Sigma}(<r_{max})$-$V_{max}$ for the eight MW and the five M31 dSphs~\cite{Hayashi17}. Note that in this plot  we have not shown the uncertainties in $V_{max}$ for the observational data  for brevity.}
\label{fig:max}
\end{figure*}

We now reconstruct a similar empirical relation between $\Sigma(<r_{max})$ and $V_{max}$
 using the three sets of simulations,  following the same methodology as K24 and compare with the same observational data.  
For this purpose, we have directly used the density profile data  for all these calculations and not relied on any of the parametric dark matter fits.  Once $r_{max}$ is obtained, we can calculate $\Sigma(<r_{max})$ from Eq.~\ref{eq:Sigma_vmax_definition}.
 We  thereby obtain the scaling relation between $\bar{\Sigma}(<r_{max})$ and 
$M (<r_{max})$ using the best-fit dark matter parameters for the three sets of simulations.   The plots showing these scaling relations  along with the resulting data for Milky Way and M31 dwarfs can be found in Fig.~\ref{fig:max}. We find that $V_{max}$ ranges from $\sim 30$ km/s  to  $\sim 40$ km/s for CDM and SIDM (with baryons).  The maximum value of  $V_{max}$ which we get for SIDM DMO only simulations is around 43 km/sec. 
We also overlay  the corresponding relations from both the dark matter models derived  in K24 on this plot. We find that the relations from the three simulations (CDM+baryons, SIDM+baryons, SIDM DMO only)  agree with the observational data on dwarfs, where the surface density is correlated with the maximum circular velocity.
We also find that  all the three  dark matter simulations analyzed in S25 agree more closely with   the cusp to core transformation model considered in K24 as opposed to  the cuspy dark matter profile based on $c-m$ relation considered in ~\cite{Moline23}.

\section{Conclusions}
\label{sec:last}
A large number of studies have shown that the dark matter halo surface density is nearly constant for a diverse  suite of galaxies  over 18 decades in blue magnitude. However, this invariance  does not hold for galaxy clusters and galaxy groups. In a previous work, we investigated the dark matter  column density (a variant of surface density) for mock catalogs from baryonified Gravity-only $\Lambda$CDM simulations. We showed that   the column density is consistent with  both a constant value as a function of halo mass as well as   a power-law dependence with halo mass~\cite{Gopika23}.

In this work we do a similar test of the constancy of dark matter halo surface density for simulated haloes of dwarf galaxy scales  at redshift $z=0$ ($M_{halo} \approx 10^{10} M_{\odot}$) that are part of the FIRE-2 project, which  were studied in great detail in S25. The simulated haloes are obtained from three different cosmological models, viz. CDM with baryons, SIDM with baryons and SIDM DMO. The SIDM simulations  were done using  a cross-section of $\sigma/m = 1~cm^2/g$.
We fit the dark matter profiles from all the simulations to a Burkert profile and the SIDM DMO simulations to Burkert, core-Einasto as well as $\alpha \beta\gamma$  (or gNFW) profile.
For the other cosmologies we used the fitted parameters for core-Einasto and 
$\alpha \beta\gamma$ profiles evaluated in S25. Using these, we calculate the dark matter surface density for Burkert profile and column density for all the other profiles.  Our plots for the halo surface density and column density can be found in Fig.~\ref{fig:g_column}, \ref{fig:cE_column}, \ref{fig:bur_column}, and \ref{fig:bur_surf}. Finally, we also estimated a relation between ${\Sigma}(< r_{max})$   and $V_{max}$ following the same methodology as  K24.

Our conclusions from the aforementioned investigations are as follows:
\begin{itemize}
\item We find that the dark matter halo surface density as well as column density are nearly constant for all the simulations using all the parametric dark matter density profiles considered.
\item The dark matter halo surface density  obtained using the Burkert profile agrees with the observational estimate of $\rho_c r_c = (141 \pm 65) M_{\odot} pc^{-2}$~\cite{Salucci19} to within $1\sigma$ for all the three cosmological models. \rthis{However, we note that Burkert profile is not a good fit for most of the CDM haloes.}
\item The ${\Sigma}(< r_{max})-V_{max}$ relation which we obtained using  the dark matter profiles agrees with the data for Milky Way and M31 dwarfs~\cite{Hayashi17} to within $1\sigma$. 
\item The ${\Sigma}(< r_{max})-V_{max}$ relation agrees more closely with the same obtained using  a cored dark matter from a cusp to core transformation model discussed in K24,   as compared to a cuspy NFW profile with $c-M$ relations from ~\cite{Moline23}. However, it does not agree with the relation obtained using a cuspy NFW profile from $c-M$ relations in ~\cite{Moline23}.
\end{itemize}

\rthis{We note that our results for surface density have been obtained from a fit to the density profiles, In a future work, we shall also calculate the surface density by considering the 2-D projection of the haloes using the full simulation data.}

\section*{Acknowledgements}
We are grateful to Maria Straight for sharing with us the data used in S25 and   for patiently answering all our queries. \rthis{We are also grateful to the anonymous referee for very useful comments and feedback on the manuscript.}

\bibliography{ref}

\appendix
\renewcommand{\thetable}{A\arabic{table}}
\setcounter{table}{0}

\section{Profile fits to CDM and SIDM}
\label{sec:appendix}
Table~\ref{tab:g-E_fit_DMO} contains halo masses, defined as $M_{halo}(<r_{200})$, and best fit parameters for core-Einasto (Eq.~\ref{eqn:coredEinasto}) and $\alpha\beta\gamma$(Eq.\ref{eqn:alphabetagamma}) profiles for the SIDM DMO density distribution. 

\begin{table*}[h]
    \caption{The best-fit parameters for the core-Einasto (Eq.~\ref{eqn:coredEinasto}) and $\alpha\beta\gamma$ (Eq.~\ref{eqn:alphabetagamma}) profile fits for SIDM DMO ($\sigma/m = 1\ {\rm cm^2\ g^{-1}}$) halos. By fixing the shape parameter at $\hat{\alpha}=0.16$ in Eq.~\ref{eqn:coredEinasto}, the fit reduces to three free parameters, where $r_c$ denotes the dark matter core radius, and $\tilde{r}_s$ and $\tilde{\rho}_s$ represent the scale radius and scale density, respectively. For the $\alpha\beta\gamma$ profile (Eq.~\ref{eqn:alphabetagamma}), fixing $\beta_{\rm s}=2.5$ and $\gamma_{\rm s}=0$ yields a three-parameter fit. In this case, $\alpha_s$ characterizes the rate at which the slope transitions between the inner and outer halo regions, while $r_s$ and $\rho_s$ denote the scale radius and scale density, respectively. The corresponding quality of fit, $Q$, is defined in Eq.~\ref{eqn:quality}.}
    \label{tab:g-E_fit_DMO}
    
    \begin{tabular*}{\textwidth}{@{\extracolsep{\fill}} lccccccccc@{}}
        \hline
        
        & & \multicolumn{4}{c}{core-Einasto} & \multicolumn{4}{c}{$\alpha\beta\gamma$}\\
        \cline{3-6} \cline{7-10}
        
        SIDM DMO halo & $M_{halo} [M_\odot]$ & $r_{\rm c} [\rm kpc]$ & $\tilde{r}_{\rm s} [\rm kpc]$ & $\tilde{\rho}_{\rm s} [M_\odot\ {\rm kpc}^{-3}]$ & $Q_{\rm cEin}$
        & $\alpha_{\rm s}$ & $r_{\rm s} [\rm kpc]$ & $\rho_{\rm s} [M_\odot\ {\rm kpc}^{-3}]$ & $Q_{\alpha\beta\gamma}$\\[2pt]

        \hline
        m10b & $6.3\times10^9$ & 0.34 & 2.94 & $3.47\times10^6$ & 0.153 & 1.18 & 0.94 & $7.72\times10^7$ & 0.205 \\
        m10c & $5.6\times10^9$ & 0.57 & 2.75 & $4.44\times10^6$ & 0.108 & 1.31 & 1.12 & $4.83\times10^7$ & 0.129 \\
        m10d & $5.3\times10^9$ & 0.53 & 1.89 & $1.14\times10^7$ & 0.127 & 2.17 & 1.19 & $3.72\times10^7$ & 0.156 \\
        m10e & $5.6\times10^9$ & 0.20 & 4.19 & $1.46\times10^6$ & 0.106 & 1.00 & 0.90 & $9.90\times10^7$ & 0.084 \\
        m10f & $5.4\times10^9$ & 0.38 & 1.81 & $1.29\times10^7$ & 0.205 & 2.59 & 0.98 & $5.88\times10^7$ & 0.075 \\
        m10h & $7.6\times10^9$ & 0.67 & 1.82 & $1.80\times10^7$ & 0.144 & 2.17 & 1.20 & $4.83\times10^7$ & 0.096 \\
        m10k & $7.1\times10^9$ & 0.77 & 1.80 & $1.87\times10^7$ & 0.142 & 2.14 & 1.27 & $4.09\times10^7$ & 0.089 \\
        m10m & $7.7\times10^9$ & 0.92 & 1.29 & $5.10\times10^7$ & 0.137 & 2.65 & 1.24 & $4.34\times10^7$ & 0.156 \\
        \hline

    \end{tabular*}
\end{table*}

\renewcommand{\thetable}{B\arabic{table}}
\setcounter{table}{0}

\clearpage
\section{Fits to Burkert and Einasto profile with all free parameters}

\label{sec:appendixB}
\rthis{ We now  examine how our  results change when  we do not fix  parameters such as $\hat{\alpha}$ (for core-Einasto), $\beta_{\rm s}$, and $\gamma_{\rm s}$ (for $\alpha\beta\gamma$ profile) a priori, but instead allow all these parameters to vary freely.
This minimization of $Q$ (with all parameters free) was done using {\tt iMinuit}, since we could not get convergence with {\tt scipy.minimize}, which we had previously used.
 The  updated parameter values   and the corresponding column densities ($S^*_{\mathrm{\alpha\beta\gamma}}$ and $S^*_{\mathrm{cEinasto}}$ can be found in Table~\ref{tab:fit_free_params} and Table~\ref{tab:fit_free_params_einasto} for $\alpha\beta\gamma$ and core-Einsato  profiles, respectively.
 We find that for both  profiles, the best-fit values for the aforementioned parameters are close to  the fixed parameter values, which we had considered earlier. For example in Table~\ref{tab:fit_free_params}, $\gamma_{\rm s}$ $\sim$ 0 and $\beta_{\rm s}$ $\sim$ 2.5,  similar to our previously fixed values. Further, in Table~\ref{tab:fit_free_params_einasto} we obtain $\hat{\alpha}$ $\sim$ 0.16 for core-Einasto profile, close to what we had fixed earlier. We now find that  $Q_{free}<Q_{fixed}$ for almost all halos due to the additional number of free parameters. }
 
 \rthis{We also  observed that a handful of  column density values  showed significant deviations from previous values. For example, the halo m10b (SIDM DMO halo),  for which the best-fit value was $\beta_{\rm s} \sim 4.0$,   differs from the previous fixed value of  2.5 by about 60\%. It also has a significantly larger scale radius $r_{\rm s} = 5.69$ kpc, which differs from the previous fitted value of 0.94 kpc by almost a  factor of six. We note however, that for this same halo, the column density for the core-Einasto profile with free $\hat{\alpha}$, is nearly the same as before. }
 
 \rthis{Similarly, for m10f (SIDM halo+baryons) the best-fit value for $\hat{\alpha}$, when fitting to the core-Einasto profile is given by  $\hat{\alpha}$ $\sim$ 0.11, which differs from the previous value of  0.16 by about 30\% and it also has a significantly smaller scale radius $\tilde{r}_{\rm s} = 0.84$~kpc,  which differs from the previous fitted value of 1.91~kpc by about 55\%. For this halo, the  column density value changes  from 67 to 112$M_\odot\,\mathrm{pc}^{-2}$. However, for this halo the column density for  $\alpha\beta\gamma$  profile is almost the same as before. }
 
 \rthis{Apart from the aforementioned exceptions, the  column density values are almost similar to the previous values for all other halos in all the simulations.}
\begin{table}[h]
\centering
\caption{\textbf{Best-fit values  of the free parameters of $\alpha\beta\gamma$ profiles with all parameters free. We also compare the new $Q$ values ($Q^{\rm \alpha\beta\gamma}_{free}$) with $Q^{\rm \alpha\beta\gamma}_{fixed}$, which we previously calculated fixing $\gamma_{\rm s}$ = 0 and $\beta_{\rm s}$ = 2.5. The new and old column densities can also be found in the last two columns.}}
\begin{tabular}{lcccccccccc}
    \hline
    $\begin{array}{c} 
        CDM\ halo \\ (+baryons)
    \end{array}$ & 
    $\begin{array}{c} 
        M_{200} \\ (M_\odot)
    \end{array}$ &
    $\begin{array}{c} 
        \gamma_{\rm s}
    \end{array}$ &
    $\begin{array}{c} 
        \beta_{\rm s}
    \end{array}$ &
    $\begin{array}{c} 
        \alpha_{\rm s}
    \end{array}$ &
    $\begin{array}{c} 
        r_{\rm s} \\ (\rm kpc)
    \end{array}$ &
    $\begin{array}{c} 
        \rho_{\rm s} \\ (M_\odot\,\mathrm{kpc}^{-3})
    \end{array}$ &
    $\begin{array}{c} 
        Q^{\mathrm{\alpha\beta\gamma}}_{free}
    \end{array}$ &
    $\begin{array}{c} 
        Q^{\mathrm{\alpha\beta\gamma}}_{fixed}
    \end{array}$ &
    $\begin{array}{c} 
        S^{\mathrm{\alpha\beta\gamma}}_{free} \\ (M_\odot\,\mathrm{pc}^{-2})
    \end{array}$ &
    $\begin{array}{c} 
        S^{\mathrm{\alpha\beta\gamma}}_{fixed} \\ (M_\odot\,\mathrm{pc}^{-2})
    \end{array}$ \\
    \hline

    m10b & $5.0\times10^9$ & 0.21 & 2.64 & 0.59 & 0.52 & $8.33\times10^{8}$ & 0.067 & 0.069 & 129 & 151\\
    m10c & $4.5\times10^9$ & 0.00 & 2.82 & 0.53 & 0.77 & $8.04\times10^{8}$ & 0.073 & 0.082 & 81 & 96\\
    m10d & $4.6\times10^9$ & 0.20 & 2.83 & 0.52 & 0.66 & $1.09\times10^{9}$ & 0.060 & 0.076 & 115 & 152\\
    m10e & $5.1\times10^9$ & 0.00 & 2.67 & 0.67 & 0.95 & $2.13\times10^{8}$ & 0.050 & 0.047 & 67 & 70\\
    m10f & $5.2\times10^9$ & 0.00 & 2.87 & 0.74 & 0.81 & $4.46\times10^{8}$ & 0.104 & 0.124 & 119 & 136\\
    m10h & $7.0\times10^9$ & 0.00 & 2.63 & 0.89 & 0.85 & $2.20\times10^{8}$ & 0.034 & 0.033 & 118 & 126\\
    m10k & $6.7\times10^9$ & 0.00 & 2.86 & 1.01 & 1.59 & $7.11\times10^{7}$ & 0.044 & 0.084 & 73 & 86\\
    m10m & $7.3\times10^9$ & 0.00 & 2.88 & 1.04 & 1.50 & $8.63\times10^{7}$ & 0.087 & 0.123 & 86 & 103\\
    
    \hline
    \multicolumn{11}{l}{SIDM halo+baryons} \\
    \hline
    m10b & $5.1\times10^9$ & 0.00 & 2.41 & 1.21 & 0.71 & $1.13\times10^{8}$ & 0.091 & 0.095 & 99 & 94\\
    m10c & $4.5\times10^9$ & 0.01 & 2.46 & 1.12 & 0.90 & $7.34\times10^{7}$ & 0.057 & 0.058 & 71 & 70\\
    m10d & $4.8\times10^9$ & 0.01 & 2.82 & 0.90 & 1.17 & $1.18\times10^{8}$ & 0.100 & 0.105 & 75 & 87\\
    m10e & $5.0\times10^9$ & 0.05 & 2.45 & 1.18 & 1.34 & $3.12\times10^{7}$ & 0.039 & 0.040 & 50 & 49\\
    m10f & $5.1\times10^9$ & 0.01 & 2.51 & 2.00 & 0.95 & $6.45\times10^{7}$ & 0.101 & 0.103 & 111 & 112\\
    m10h & $7.0\times10^9$ & 0.00 & 2.52 & 1.59 & 1.22 & $5.10\times10^{7}$ & 0.040 & 0.039 & 94 & 94\\
    m10k & $6.6\times10^9$ & 0.01 & 2.72 & 1.60 & 1.82 & $2.80\times10^{7}$ & 0.052 & 0.092 & 68 & 75\\
    m10m & $7.3\times10^9$ & 0.01 & 2.83 & 1.21 & 1.65 & $5.32\times10^{7}$ & 0.094 & 0.120 & 78 & 91\\
    \hline
    \multicolumn{11}{l}{SIDM DMO halo} \\
    \hline
    m10b & $6.3\times10^9$ & 0.14 & 4.08 & 0.51 & 5.69 & $1.74\times10^{8}$ & 0.163 & 0.205 & 22 & 81 \\
    m10c & $5.6\times10^9$ & 0.00 & 3.03 & 0.77 & 1.75 & $9.60\times10^{7}$ & 0.107 & 0.130 & 53 & 68 \\
    m10d & $5.3\times10^9$ & 0.01 & 2.87 & 1.06 & 1.37 & $7.91\times10^{7}$ & 0.112 & 0.135 & 75 & 85 \\
    m10e & $5.6\times10^9$ & 0.00 & 2.54 & 0.94 & 0.90 & $1.14\times10^{8}$ & 0.085 & 0.084 & 78 & 79 \\
    m10f & $5.4\times10^9$ & 0.00 & 2.56 & 2.00 & 1.01 & $6.65\times10^{7}$ & 0.077 & 0.092 & 117 & 122 \\
    m10h & $7.6\times10^9$ & 0.00 & 2.66 & 1.57 & 1.35 & $5.66\times10^{7}$ & 0.077 & 0.099 & 103 & 111 \\
    m10k & $7.1\times10^9$ & 0.03 & 2.67 & 1.60 & 1.46 & $4.57\times10^{7}$ & 0.065 & 0.091 & 92 & 99 \\
    m10m & $7.7\times10^9$ & 0.00 & 2.88 & 1.42 & 1.64 & $5.68\times10^{7}$ & 0.092 & 0.170 & 97 & 114 \\
    \hline
\end{tabular}
\label{tab:fit_free_params}
\end{table}

\begin{table}[h]
\centering
\caption{\rthis{This table is similar to  Table~\ref{tab:fit_free_params}, where we fit for the  core-Einasto profile, using Eq.~\ref{eqn:coredEinasto} without fixing $\hat{\alpha}$. Here also $Q^{\rm cEinasto}_{free}$ $<$ $Q^{\rm cEinasto}_{fixed}$ and $\hat{\alpha}= 0.16$, which we had previously fixed.}}
\begin{tabular}{lccccccccc}
    \hline
    $\begin{array}{c} 
        CDM\ halo \\ (+baryons)
    \end{array}$ & 
    $\begin{array}{c} 
        M_{200} \\ (M_\odot)
    \end{array}$ &
    $\begin{array}{c} 
        \hat{\alpha}
    \end{array}$ &
    $\begin{array}{c} 
        r_{\rm c} \\ (\rm kpc)
    \end{array}$ &
    $\begin{array}{c} 
        \tilde{r}_{\rm s} \\ (\rm kpc)
    \end{array}$ &
    $\begin{array}{c} 
        \tilde{\rho}_{\rm s} \\ (M_\odot\,\mathrm{kpc}^{-3})
    \end{array}$ &
    $\begin{array}{c} 
        Q^{\mathrm{cEinasto}}_{free}
    \end{array}$ &
    $\begin{array}{c} 
        Q^{\mathrm{cEinasto}}_{fixed}
    \end{array}$ &
    $\begin{array}{c} 
        S^{\mathrm{cEinasto}}_{free} \\ (M_\odot\,\mathrm{pc}^{-2})
    \end{array}$ &
    $\begin{array}{c} 
        S^{\mathrm{cEinasto}}_{fixed} \\ (M_\odot\,\mathrm{pc}^{-2})
    \end{array}$ \\
    \hline

    m10b & $5.0\times10^9$ & 0.10 & 0.10 & 2.70  & $3.00\times10^{6}$ & 0.069 & 0.107 & 39 & 28\\
    m10c & $4.5\times10^9$ & 0.12 & 0.14 & 3.69  & $1.52\times10^{6}$ & 0.077 & 0.091 & 27 & 20\\
    m10d & $4.6\times10^9$ & 0.13 & 0.03 & 3.39  & $1.85\times10^{6}$ & 0.056 & 0.075 & 32 & 30\\
    m10e & $5.1\times10^9$ & 0.10 & 0.32 & 3.23  & $2.25\times10^{6}$ & 0.054 & 0.087 & 30 & 20\\
    m10f & $5.2\times10^9$ & 0.19 & 0.00 & 3.56  & $2.07\times10^{6}$ & 0.135 & 0.113 & 37 & 53\\
    m10h & $7.0\times10^9$ & 0.11 & 0.33 & 1.94  & $1.01\times10^{7}$ & 0.069 & 0.100 & 69 & 45\\
    m10k & $6.7\times10^9$ & 0.14 & 0.71 & 1.78  & $1.56\times10^{7}$ & 0.077 & 0.085 & 65 & 57\\
    m10m & $7.3\times10^9$ & 0.16 & 0.58 & 2.08  & $1.13\times10^{7}$ & 0.102 & 0.102 & 66 & 69\\
    
    \hline
    \multicolumn{10}{l}{SIDM halo+baryons} \\
    \hline
    m10b & $5.1\times10^9$ & 0.10 & 0.30 & 1.79  & $8.69\times10^{6}$ & 0.116 & 0.144 & 56 & 32\\
    m10c & $4.5\times10^9$ & 0.11 & 0.40 & 2.24  & $5.15\times10^{6}$ & 0.095 & 0.120 & 40 & 25\\
    m10d & $4.8\times10^9$ & 0.13 & 0.43 & 1.75  & $1.06\times10^{7}$ & 0.110 & 0.120 & 57 & 43\\
    m10e & $5.0\times10^9$ & 0.11 & 0.70 & 2.67  & $4.06\times10^{6}$ & 0.076 & 0.094 & 32 & 22\\
    m10f & $5.1\times10^9$ & 0.11 & 0.58 & 0.71  & $1.06\times10^{8}$ & 0.175 & 0.197 & 112 & 67\\
    m10h & $7.0\times10^9$ & 0.11 & 0.77 & 1.18  & $4.17\times10^{7}$ & 0.129 & 0.149 & 86 & 57\\
    m10k & $6.6\times10^9$ & 0.14 & 1.29 & 1.21  & $5.43\times10^{7}$ & 0.141 & 0.147 & 73 & 67\\
    m10m & $7.3\times10^9$ & 0.15 & 0.84 & 1.61  & $2.31\times10^{7}$ & 0.121 & 0.124 & 74 & 71\\
    \hline
    \multicolumn{10}{l}{SIDM DMO halo} \\
    \hline
    m10b & $6.3\times10^9$ & 0.19 & 0.21 & 3.94  & $1.79\times10^{6}$ & 0.151 & 0.154 & 31 & 39\\
    m10c & $5.6\times10^9$ & 0.16 & 0.46 & 2.93  & $3.56\times10^{6}$ & 0.098 & 0.097 & 37 & 39\\
    m10d & $5.3\times10^9$ & 0.15 & 0.54 & 1.77  & $1.29\times10^{7}$ & 0.126 & 0.127 & 62 & 60\\
    m10e & $5.6\times10^9$ & 0.11 & 0.36 & 2.60  & $4.08\times10^{6}$ & 0.090 & 0.106 & 40 & 28\\
    m10f & $5.4\times10^9$ & 0.11 & 0.53 & 0.84  & $7.51\times10^{7}$ & 0.185 & 0.205 & 114 & 74\\
    m10h & $7.6\times10^9$ & 0.13 & 0.78 & 1.24  & $4.49\times10^{7}$ & 0.136 & 0.145 & 97 & 80\\
    m10k & $7.1\times10^9$ & 0.13 & 0.90 & 1.07  & $6.37\times10^{7}$ & 0.131 & 0.142 & 96 & 75\\
    m10m & $7.7\times10^9$ & 0.16 & 0.86 & 1.37  & $4.12\times10^{7}$ & 0.139 & 0.138 & 96 & 101\\
    \hline
\end{tabular}
\label{tab:fit_free_params_einasto}
\end{table}

\label{lastpage}


\end{document}